\documentclass{PoS}
\newcommand{\gtsim}{\mbox{{\raisebox{-0.4ex}{$\stackrel{>}{{\scriptstyle\sim}}
$}}}}
\newcommand{\ltsim}{\mbox{{\raisebox{-0.4ex}{$\stackrel{<}{{\scriptstyle\sim}}
$}}}}
\title{Multi-wavelength surveys towards the SKA pathfinders}

\ShortTitle{Multi-wavelength surveys}

\author{\speaker{Matt J.~Jarvis}\thanks{RCUK Fellow}\\
        Centre for Astrophysics Research, STRI, University of Hertfordshire, Hatfield, AL10 9AB\\
        E-mail: \email{M.J.Jarvis@herts.ac.uk}}

\abstract{In these proceedings I discuss various extragalactic surveys which will be undertaken over the next few years and which will be complementary to any H{\sc i} and/or continuum surveys with the SKA-pathfinder telescopes. I concentrate on the near-infrared public surveys which will be undertaken with the Visible and Infrared Survey Telscope for Astronomy (VISTA), and in particular the VISTA Deep Extragalactic Observations (VIDEO) survey which will provide the ideal data set to combine with any deep SKA-pathfinder observations of the extragalactic sky. After highlighting the links that the SKA pathfinders have with the various VISTA surveys, I briefly describe an approved large area survey to be carried out with the {\sl Herschel Space Observatory} which has a large scientific overlap with the SKA pathfinder telescopes.}

\FullConference{From planets to dark energy: the modern radio universe\\
		 October 1-5 2007\\
		 University of Manchester, Manchester, UK}

\begin{document}

\section{Introduction}

The Square Kilometre Array promises to be one of the most influential telescopes of any era in astronomy. In the area of galaxy formation and evolution, it will be able to map out the neutral Hydrogen content of galaxies up to and possibly beyond $z \sim 1.5$. However, it will be insensitive to the underlysing stellar populations and various physical mechanisms that are associated with galaxies which do not emit at radio wavelengths. Thus, even in the age of the SKA complementary multi-wavelength surveys will still be paramount to further our understanding of the Universe. Furthermore, the various SKA pathfinder telescopes will also benefit greatly from multi-wavelength surveys which are being or will be carried out concurrently with the deep radio surveys. 

Such surveys are already underway or being planned, and many more will undoubtedly be initiated in the coming years. In these proceedings I focus on the near-infrared surveys which will be carried out by the ESO-Visible and Infrared Survey Telescope for Astronomy (VISTA), to begin operation in 2008 at Paranal, Chile, in particular the second deepest tier of the extragalactic public surveys to be undertaken on this telescope, namely the VISTA Deep Extragalactic Observations (VIDEO) survey.

I also highlight a recently approved large-area survey to be conducted with the {\sl Herschel Space Observatory} which I believe will be of benefit and benefit from observations with the various SKA pathfinder telescopes.

\section{Extragalactic near-infrared surveys with VISTA}\label{sec:vistasurveys}

VISTA will conduct several public surveys at near-infrared wavelengths ($0.8 - 2.4\mu$m). There are currently six approved surveys, I briefly discuss three of the surveys most relevant for extragalactic studies with the SKA pathfinders and in section~\ref{sec:VIDEO} I discuss in more detail the VIDEO survey\footnote{Further details of all of the public surveys to be conducted on both the VISTA and VST can be found in Arnaboldi et al. 2007.}: 

\begin{itemize}

\item Ultra-VISTA is the deepest tier of the extragalactic surveys. It will survey the COSMOS field with three separate strategies. The ultra-deep survey covers 0.73~deg$^{2}$ to AB magnitudes of $Y=26.7$, $J=26.6$, $H=26.1$ and $K_{s}=25.6$, with the aim of detecting galaxies within the epoch of reionisation at $z>6$. The narrow-band survey is expected to find $\sim 30$ Ly$\alpha$ emitters at $z \sim 8.8$ to a depth of NB$_{\rm AB}=24.1$ and the wide survey will complete the coverage of the full 1.5~deg$^{2}$ COSMOS field, to depths of  $Y=25.7$, $J=25.5$, $H=25.1$ and $K_{s}=24.5$ (all AB).
With reference to the science aims of the SKA and its pathfinders, the Ultra-VISTA survey does not cover enough area to be fully utilised with a radio deep field from these facilities which have relatively large fields-of-view.

\item The VISTA Kilo-degree Infrared Galaxy (VIKING) survey aims to survey two stripes at high galactic latitude in five near-infared filters. The areas have been chosen to overlap with the regions of sky covered by the 2df galaxy redshift survey (Colless et al. 2001) and the optical Kilo-Degree Survey (KIDS) to be conducted with the VLT Survey Telescope, thus providing both spectroscopic redshift up to $z\sim 0.3$ and photometric redshifts up to and above $z\sim 1$ with typical uncertainties of $\Delta z/(1+z) \sim 0.1$ from this 9-band survey. The science aims are heavily based on this photometric redshift accuracy for cosmological studies to constrain the dark energy component in the Universe, gain more accurate mass density measurement from weak lensing, find $z>7$ quasars and trace galaxy evolution from $z\sim 1$ to the present day. These regions are also the subject of the Herschel large-area survey (see section~\ref{sec:H1K}).
It is apparent that any $\sim 1000$~deg$^{2}$ survey from the SKA pathfinders should target these regions to maximise the scientific productivity of both continuum and H{\sc i} surveys.

\item The VISTA Hemisphere Survey (VHS) aims to survey the {\it rest} of the southern sky which is not being covered by the other VISTA surveys. Although much shallower than the other surveys it will provide data covering $\sim 18000$~deg$^{2}$ to a depth of at least $K_{\rm s} \sim 19.8$ and $J\sim 20.9$.
The majority of this area will alslo be covered by various surveys conducted at optical wavelengths, namely the Dark Energy Survey (DES; https://www.darkenergysurvey.org/) and the VST-ATLAS survey, in addition to the various galactic plane studies. Such a survey has wide-ranging scientific goals and its complementarity to any large area SKA pathfinder survey is obvious. If the SKA pathfinders progress as imagined toward the full SKA then it is plausible that H{\sc i} redshifts could be obtained for a significant fraction of the VHS galaxies, allowing detailed investigations between the H{\sc i} and the stellar properties of the galaxies.

\end{itemize}
\includegraphics[width=15.5cm]{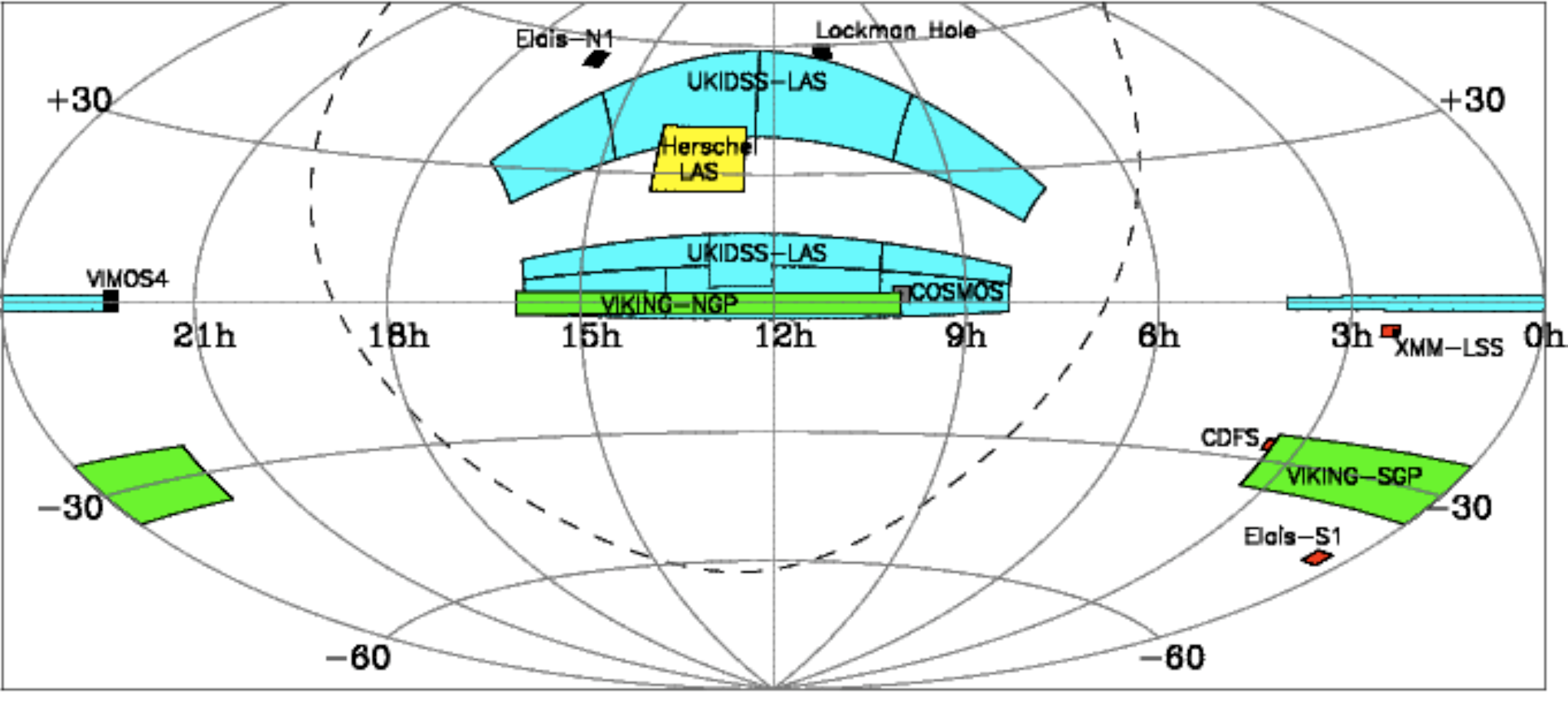}
{\textsf {\bf Figure~1.} Overview of the various VISTA surveys described in the text along with the regions to be observed with the UK Infrared Deep Sky Survey (UKIDSS; www.ukidss.org) and the Herschel large-area survey (Herschel-LAS). (Best viewed in colour.)}


\section{VISTA Deep Extragalactic Observations (VIDEO) survey}\label{sec:VIDEO}

In this section I describe the VISTA Deep Extragalactic Observations (VIDEO) survey. VIDEO aims to obtain both deep and wide near-infrared observations over well-studied contiguous fields covering $3-4.5$~deg$^{2}$ each. The fields chosen for this are derived from the equatorial and southern fields of the Spitzer Wide-area Infrared Extragalactic Survey (SWIRE; Lonsdale et al. 2003), namely $\sim 4.5$~deg$^{2}$ within the Elais-S1 region, $\sim 4.5$~deg$^{2}$ within the XMM-Newton Large Scale Structure (XMM-LSS) survey and 3~deg$^{2}$ around the Chanda Deep Field South (CDFS). 

The VIDEO survey is planned to reach the following $5\sigma$ (2~arcsec point source) depths; $Z=25.7$, $Y=24.6$, $J=24.5$, $H=24.0$ and $K_{\rm s}=23.5$ (all AB magnitudes), and will use just over 200~nights of observing time over the next five years. In terms of the galaxies which will be detected in the VIDEO survey, these depths correspond to an $L^{\star}$ elliptical galaxy up to $z\sim4$ and $0.1~L^{\star}$ galaxy up to $z\sim1$, thus it sits naturally between the Ultra-VISTA and VIKING surveys.

This combination of depths and area ensures that with the VIDEO survey we will be able to trace galaxy formation and evolution over the majority of cosmic history and over all environmental densities. Some of the main science goals of VIDEO will be; 

\begin{itemize}

\item to trace the evolution of galaxies from the earliest epochs until the present day. The depth and breadth of the VIDEO survey will enable galaxy properties to be measured as a function of environmental density,  allowing the detailed study of how the the environmental richness may affect the properties of the galaxies. Furthermore, the combination of VIDEO data with the wealth of multi-wavelength data over these regions of sky, will allow the star-formation rate and AGN activity to be be measured and linked to the stellar mass in the galaxies.

\item to measure the clustering of the most massive galaxies at $z>5$ where, extrapolating from the results of McLure et al. (2006), we expect to detect around 300 galaxies with $M>10^{11}$~M$_{\odot}$ and around 150 galaxies at $z>6$.

\item to trace the evolution of galaxy clusters from the formation epoch until the present day (see e.g. van Breuekelen et al. 2006). The depth of VIDEO ensures that we will be able to trace the bright end of the cluster luminosity function to $z\sim 3$, while its area should provide a sample of $\sim 70$ clusters with $M>10^{14}$M$_{\odot}$ at $z>1$, with around 15 of these expected to be at $z>1.5$.

\item to quantify the accretion activity over the history of the Universe. The depth and area, along with the filter combination will allow the detection of the highest redshift quasars, and this place the first constraints on the quasar luminosity function at $z>6.5$, when combined with  wider and shallower surveys, such as VIKING. Furthermore, VIDEO data can be combined with Spitzer and Herschel data to place constraints on both the obscured AGN and star-forming galaxies, all the way out to $z\sim6$ if AGN host galaxies have typical luminosities of $2-3~L^{\star}$ (e.g. Jarvis et al. 2001; Mart\'inez-Sansigre et al. 2006).

\end{itemize} 

The VIDEO survey will therefore provide the ideal data to be combined with any deep extragalactic observations with the SKA-pathfinder telescopes in both continuum and H{\sc i}. Crucially, VIDEO will have the depth to detect low-mass galaxies which would dominate the $\ltsim1$ SKA H{\sc i} surveys, thus providing the constraints on the stellar populations in these galaxies.

\section{Large Area Herschel Survey}\label{sec:H1K}

Although I have highlighted the near-infrared surveys that will be undertaken with the VISTA telescope it is also important to consider what data at other wavelengths will be available on the timescale of the SKA-pathfinder telescopes. The list is obviously extensive and spans the whole of the electromagnetic spectrum, therefore here I only highlight a large-area survey recently approved as a {\sl Herschel Space Observatory} Open Time Key Project.

From studies of the extragalactic background light, it has become clear that approximately half of the energy emitted by objects in the Universe has been absorbed  and re-radiated at mid- to far-infrared wavelengths (i.e around 60--500$\mu$m). However, we know very little about the sources responsible for this re-raditaed emission, even in the local Universe, with most of the information based on the IRAS survey carried out in the 1980s.

This will now be surpassed with the {\sl Herschel Space Observatory} which will operate at 70--500$\mu$m.
The {\sl Herschel Space  Observatory} will conduct a large area survey as one of its Open Time Key Projects. The survey will be carried out over a number of regions within the VIKING/KIDS surveys regions (see Fig.~1) and also a northern region which encompasses the Coma Cluster. The survey will use five photometric bands to conduct the first wide-area survey at sub-millimetre wavelengths.

Some of the main science goals are to; (i) measure the luminosity functions and dust along the Hubble Sequence, down to dust masses of $\sim 10^{4.5}$~M$_{\odot}$, and also determine the amount of starlight which is extinguished by dust with a study of $\gtsim 10^{4}$ galaxies. (ii) investigate the environmental dependence of star formation and downsizing, very similar in goals to the VIDEO survey but at $z\ltsim 0.2$, and trace the evolution of dust and obscured star formation over $\sim 3$~billion years. (iii) find rare gravitational lenses and study galaxy mass profiles from $z\sim1.5$ to the present day. (iv) investigate the relationship between star formation and accretion activity and determine whether the AGN phenomena is intimately linked to obscured star-formation.
The science possible with the SKA pathfinders are extremely complementary to this survey, and it will be possible to link the gas with the dust emission from the galaxies detected in both surveys.

\section{Acknowledgements}

I would like to thank the various PIs of the surveys which I discuss for allowing me to steal figures and text from many different proposals, and also for actually leading the surveys. Thanks also to Simon Dye who provided the code to generate figure~1.

\end{document}